\documentclass[12pt]{article}

\usepackage{scicite}

\usepackage{times}

\usepackage{graphicx}
\usepackage{dcolumn}
\usepackage{epstopdf}
\usepackage{bm}
\usepackage{color}
\usepackage{ulem}

\def\pr#1#2#3{{\it Phys.~Rev.}~{\bf #1},\ #2\ (#3)}
\def\pra#1#2#3{{\it Phys.~Rev.~A}~{\bf #1},\ #2\ (#3)}

\def\prl#1#2#3{{\it Phys.~Rev.~Lett.}~{\bf #1},\ #2\ (#3)}
\def\sci#1#2#3{{\it Science}~{\bf #1},\ #2\ (#3)}
\def\nat#1#2#3{{\it Nature}~{\bf #1},\ #2\ (#3)}
\def\natphys#1#2#3{{\it Nat.~Phys.}~{\bf #1},\ #2\ (#3)}

\newcommand{\etal}{{\it et al.}}

\newenvironment{sciabstract}{%
\begin{quote} \bf}
{\end{quote}}

\newcounter{lastnote}
\newenvironment{scilastnote}{%
\setcounter{lastnote}{\value{enumiv}}%
\addtocounter{lastnote}{+1}%
\begin{list}%
{\arabic{lastnote}.}
{\setlength{\leftmargin}{.22in}}
{\setlength{\labelsep}{.5em}}}
{\end{list}}

\title{Production of trilobite Rydberg molecule dimers with thousand-Debye permanent electric dipole moments}

\author
{D. Booth,$^{1}$ S. T. Rittenhouse,$^{2}$ J. Yang,$^{1}$ H. R. Sadeghpour,$^{3\ast}$ J.P. Shaffer$^{1}$\\
\\
\normalsize{$^{1}$Homer L. Dodge Department of Physics and Astronomy, University of Oklahoma,}\\
\normalsize{440 W. Brooks St., Norman, OK 73019, USA}\\
\normalsize{$^{2}$Deptartment of Physics and Astronomy, Western Washington University,}\\
\normalsize{516 High St., Bellingham, WA 98225, USA}\\
\normalsize{$^{3}$ITAMP, Harvard-Smithsonian Center for Astrophysics, 60 Garden St., Cambridge, MA 02138, USA}\\
\\
\normalsize{$^\ast$To whom correspondence should be addressed; E-mail:  hrs@cfa.harvard.edu.}
}

\date{}

\begin{document}


\baselineskip24pt


\maketitle


\begin{sciabstract}
We observe that when an ultracold ground state cesium (Cs) atom becomes bound within the electronic cloud of an extended Cs electronic orbit, ultralong-range molecules with giant (kilo-Debye) permanent electric dipole moments form. Large molecular permanent electric dipole moments are challenging to experimentally realize. Meeting this challenge has garnered significant interest because permanent electric dipole moments are important for understanding symmetry breaking in molecular physics, control of chemical reactions and realization of strongly correlated many-body quantum systems. { These new hybrid class of `trilobite' molecules are predominated with degenerate Rydberg manifolds, making them difficult to produce with conventional optical association.} Their behavior is quantitatively reproduced with detailed coupled-channel calculations. 
\end{sciabstract}


Dipole moments are fundamentally important for control of chemical reactions and dynamics \cite{onw10,cdk09}, precision spectroscopy, realization of strongly-correlated many-body gases exhibiting novel quantum phase transitions \cite{weimer10}, quantum information processing \cite{lukin01}, and tests of fundamental symmetries \cite{cdk09}. Permanent electric dipole moments (PEDM) in molecules are a manifestation of symmetry-breaking. They form in quantum systems by charge separation and mixing of opposite parity eigenstates. Homonuclear molecules are therefore not expected to possess PEDM \cite{klemp97}.

It was shown in Ref. \cite{li11} that ultralong-range rubidium (Rb) Rydberg dimers correlating to $nS + 5S$ molecular asymptotes, with $n$ the Rydberg atom's principal quantum number, possess sizable PEDMs, $\sim 1\,$Debye, and a linear Stark map.  It was also predicted that due to the smaller noninteger component of the Cs $S$-state quantum defect, Cs Rydberg molecules correlating to $nS+6S$ asymptotes may have PEDM as large as $\sim 15$ Debye. In contrast to states with angular momentum quantum numbers $l \leq2$ \cite{gre06,bend09,bluetrilo,Eyler13,Pfau14,Raithel14}, which are most easily addressed by laser excitation, the original theory of ultralong-range Rydberg molecules \cite{gds00} predicted huge kilo-Debye PEDM would form in the excitation of a completely degenerate Rydberg manifold. Cs atoms, because of their energy level structure and ground state electron scattering properties are ideal for observing a new species of ultralong-range Rydberg molecule that is a hybrid of the low and high $l$-type molecules. This class of Cs ultralong-range Rydberg molecules possess giant PEDMs but are accessible via conventional two-photon laser excitation.

\begin{figure}[t!]
\begin{center}
\resizebox{1.0\columnwidth}{!}{\includegraphics{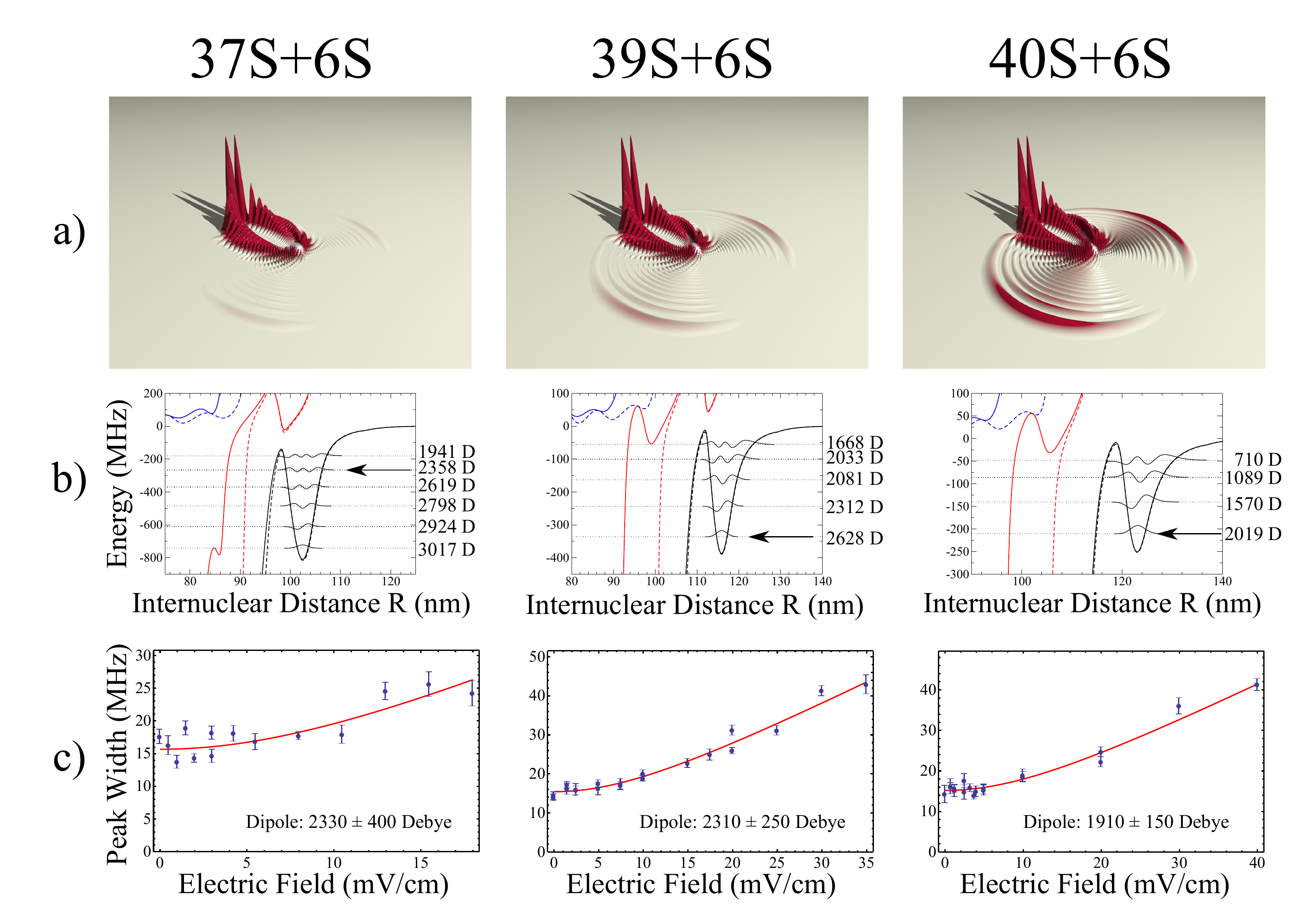}}
\caption{\label{fig1}  a) The electron probability distribution, $|\Psi(R_e;\mathbf{r}) |^2$ in Eq. (4),  for the molecular states to the red of Cs($n=37S$, $39S$, and $40S$) Rydberg lines, each marked in (b) with an arrow, in cylindrical coordinates. The Rydberg ion is located near the center of each plot and the Rydberg electron is localized near the location of the ground state perturber. The distributions are shown at the equilibrium separations at, $R_e=102.4$, $115.9$, and $123.0\,$nm, for the corresponding BO PECs, given in (b). The vibrational wave functions in the outermost wells are indicated. The BO PECs (black, red and blue curves) correlate asymptotically to the $nS+6S$, the $(n-4)F+6S$, and the  $(n-4)G+6S$ states. The dashed (solid) curves have the $M_J=0$ ($M_J=\pm 1$) projection symmetry. The vibrationally averaged PEDM are also shown for each $v$.
c) Linewidths as a function of $F_a$ for the arrow marked states in (b). The error bars for the PEDM are determined as the background field and two-photon laser linewidth are varied within the measurement error. The error bars on the linewidth data are the statistical error of the linewidth fit.}
\end{center}
\end{figure}

\begin{figure}[t!]
\resizebox{1.0\columnwidth}{!}{\includegraphics{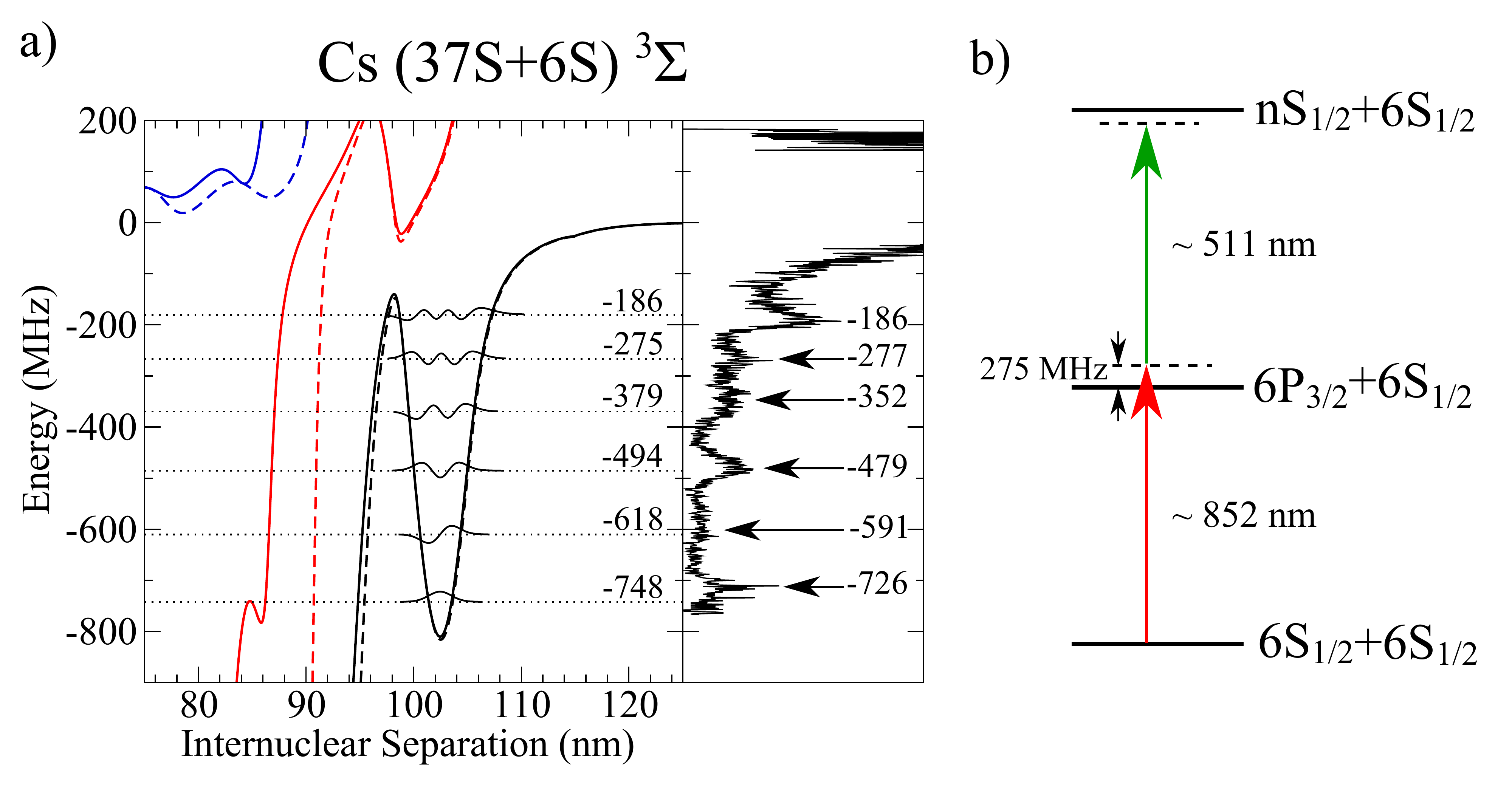}}
\caption{\label{fig2}  a) A comparison between the calculated vibrational levels in the outermost PEC well, superposed with the associated wave functions (left) and observed spectra (right), for states correlating to the $37S+6S$ limit. Even parity ($v=0,\ 2,\ ...$) vibrational levels have stronger signals because the de Broglie wavelength of the ground state wave function, $\lambda_{dB} \sim 35 \mathrm{\:nm}$, is much longer than the width of the outermost potential well.  b) Level diagram for the two-photon excitation scheme.}
\end{figure}

\begin{figure}[t!]
\resizebox{1.0\columnwidth}{!}{\includegraphics{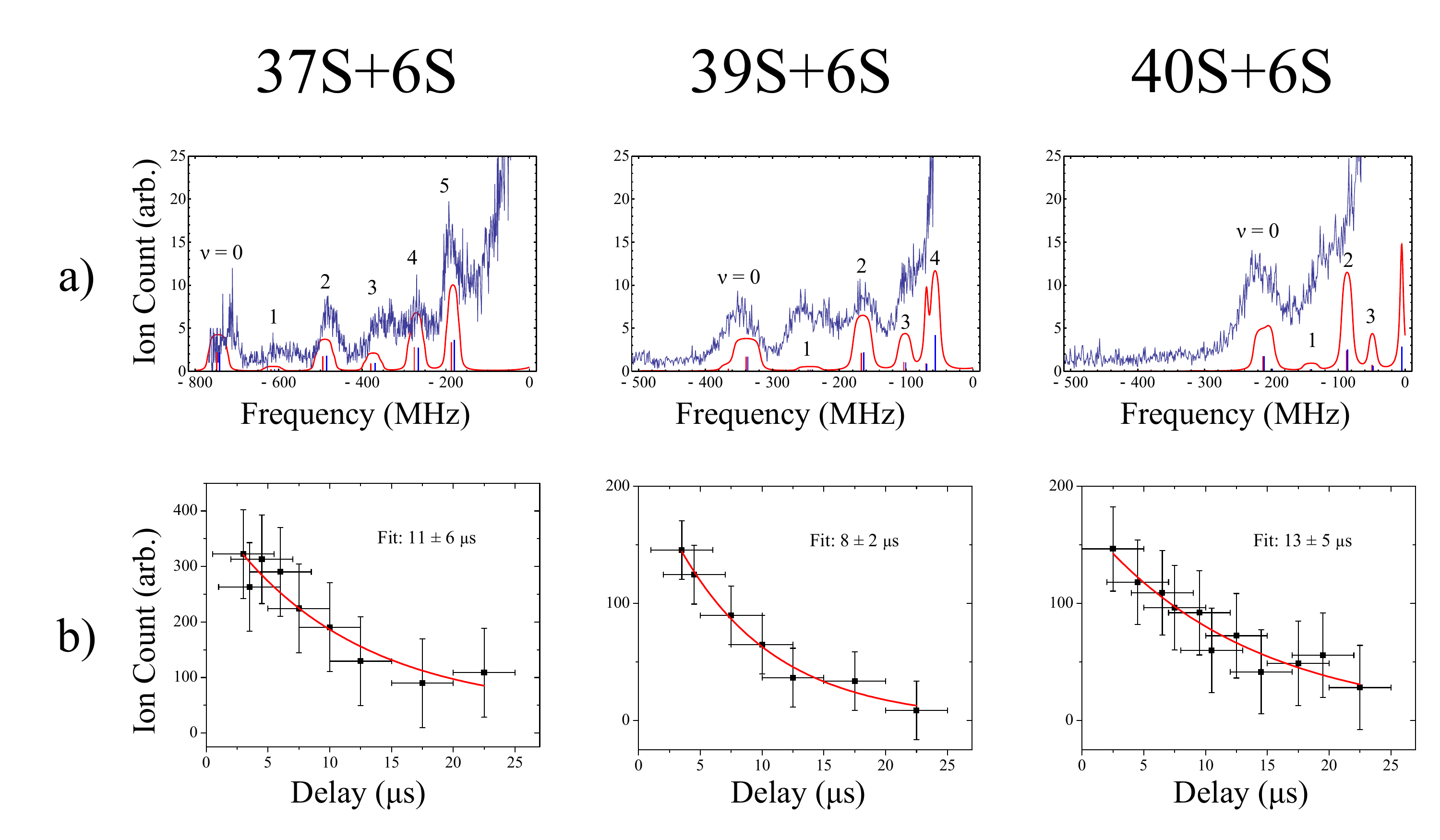}}
\caption{\label{fig3} a) A comparison between calculated spectra (red) and experimental spectra (blue) for the states correlating to the $37S + 6S$, $39S + 6S$ and $40S + 6S$ dissociation limits. The centroids of the vibrational levels are shown as sticks underneath the calculated spectra. Bound states corresponding to the $M_J=0$ and $M_J=\pm 1$ projections are indicated with red and blue sticks, respectively.
b) Ion counts for the $v=4$ level in the PEC correlating to the $37S+6S$ limit, and $v=0$ in PECs correlating to the $39S+6S$ and $40S+6S$ limits, as a function of the delay between excitation and ionization. The uncertainties in the delay time are due to the width of the $5 \mathrm{\:\mu s}$ laser pulses used. The vertical error bars are the statistical error in the ion counts.}
\end{figure}

We demonstrate that for Cs atoms, the peculiarly small noninteger fraction of the $S$-state quantum defect strongly admixes the $(n-4) l> 2$ degenerate electronic manifold with spherically symmetric nondegenerate $nS$ states to form ultralong-range Rydberg molecules with kilo-Debye PEDM, which are spectroscopically accessible. The Rydberg electron probability distributions for the observed Cs($nS-6S$) $^3\Sigma$ Rydberg molecules are predominantly of the `trilobite' type, Fig.~\ref{fig1}a, where, in this notation, the label in brackets denotes the separated atom limit. The fractional mixing of high angular momentum states can be as large as $90\%$. This contrasts with the $0.01\%$ admixture of hydrogenic state character that occurs in Rb Rydberg molecules \cite{li11}. The large admixture of nearly degenerate electronic manifold localizes the electron density on the Cs($6S$) perturber, Fig.~\ref{fig1}a. We measure the PEDM by monitoring how the molecular Rydberg lines broaden when subjected to $F \sim 30$ mV/cm external electric field, Fig.~\ref{fig1}c. Quantitative calculations of potential energy curves (PEC) with complicated non-adiabatic avoided crossings, vibrational energy levels and PEDMs corroborate the observations, Fig.~\ref{fig1}b, Fig.~\ref{fig2}a, and Fig.~\ref{fig3}a.

Excitation into Rydberg states in a quantum gas has the potential for probing many-body effects with high precision and creating exotic states of matter. A recent observation of Rydberg electron orbital excitation to sizes comparable to or exceeding the extent of a Bose-Einstein condensate (BEC) heralds possibilities for charged impurity research with extremely low mass ratios \cite{balewski13}. Due to Rydberg blockade \cite{lukin00}, only one Rydberg atom is excited in the BEC and single impurity studies can be conducted. Another important analogy between the formation of ultralong-range Rydberg molecules and impurity interaction in solids is localization \cite{and58}. The formation of ultralong-range Rydberg molecules is manifestly through multiple scattering of electrons from perturbers, leading to localization of the electronic wave packet. The Cs states produced in this work are precursors to fascinating states in materials where the electron is strongly localized at the position of several ground state atoms. Such states will have exotic properties as they can involve dipolar and spin degrees of freedom as well as interactions between Rydberg atoms, if more than one Rydberg atom is present.

The interaction between a Rydberg electron and a ground state perturber within the Rydberg atom can be described by a Fermi contact interaction
with energy dependent scattering lengths \cite{fermi34,omont77}. For Cs, there is a significant $p$-wave spin-orbit splitting, manifested in the $^3P_J$ electron-cesium scattering phase shifts that must be considered.  To calculate the molecular states for the Cs dimer, we diagonalize the electronic Hamiltonian that results from the electron atom interaction for a range of $R$ using a large basis set of Rydberg electron orbital wave functions. The calculation yields a set of Born-Oppenhiemer (BO) PEC, $U(R)$, and their corresponding Rydberg electron wave functions, $\Psi(R;\mathbf{r})$. Examples of these PECs and electron density distributions calculated from the $\Psi(R;\mathbf{r})$ are shown in Fig.~\ref{fig1}a-b. The resulting set of coupled Schr\"{o}dinger equations are solved directly to extract the vibrational states and the full spectrum of Rydberg molecular states. The depth of the BO PECs with respect to the $nS+6S$ molecular asymptotes show extreme sensitivity to the value of the zero-energy $s$-wave scattering length. This results from the fact that the depth of the PECs with respect to the hydrogenic manifold is approximately proportional to the scattering length.  Since the hydrogenic manifold lies several GHz above the Cs Rydberg $S$-state, a $1\%$ variation in the electron-ground state atom $s$-wave scattering length results in $\sim 100\,$MHz change in the PEC depth. We used this sensitivity to adjust the value of the $s$-wave scattering length so that the lowest vibrational level in the outer well of the PEC correlating to $40S+6S$ is in agreement with the experimentally observed resonance peak. The $s$-wave scattering length obtained was $-21.266\,$a$_0$. This value is $2\%$ smaller than the theoretically calculated value of $-21.7\,$a$_0$ used in \cite{bah01,bluetrilo}.

The $p$-wave electron-perturber scattering creates a set of narrow avoided crossings in the BO PECs, evident in Fig.~\ref{fig1}b. These features correspond to metastable $p$-wave Cs$^{-}$ states \cite{hgs02}. The PEC crossings can have a significant impact on the overall behavior of the molecular PECs- and the resulting vibrational states- particularly for those states lying nearby in energy. Far from the crossings, the PECs are dominated by $s$-wave scattering. We focus on states in the outer-most PEC wells, seen in Fig.~\ref{fig1}b, where the $p$-wave scattering only produces a small energy shift and the spin-orbit splitting between the $M_{J}=\pm1$ and $0$ bound-state energies is smaller than the experimental spectral resolution of $~ 3\,$MHz.

The experiment was performed in a far-off resonance trap (FORT). The crossed FORT was loaded to a peak density of $5 \times 10^{13} \mathrm{\:cm}^{-3}$ at a temperature of $40 \mathrm{\:\mu K}$. A two-photon excitation was used to photoassociate the molecules, Fig.~\ref{fig2}b. The molecules were ionized using the FORT beams. The ions were detected with a microchannel plate (MCP) detector. The molecular spectra were acquired by counting $\mathrm{Cs}^+$ and $\mathrm{Cs_2}^+$ ions arriving at the MCP as a function of excitation laser frequency. No $\mathrm{Cs_2}^+$ ions were detected.

Molecular spectra were acquired to the red of the $n=37$, $39$, and $40S_{1/2}$ atomic states. Three spectral absorption lines were selected for Stark shift measurements: the excited vibrational level at $\sim -277 \mathrm{\:MHz}$, $v=4$, in the PEC correlating to the $37S+6S$ limit, and the ground vibrational levels, $v=0$, in the PECs correlating to the $39S+6S$ and $40S+6S$ limits. These states are indicated in Fig.~\ref{fig1}b with arrows and their ion yield spectra are shown in Fig.~\ref{fig2}a and Fig.~\ref{fig3}a. For the Stark shift measurements, external electric field, $F_a$, scans were performed on the spectral lines to determine their positions and widths. Since the electric field plates can only apply electric fields normal to the plates in our apparatus, a constant horizontal background electric field of $F_h \sim 15 \mathrm{\:mV}\:\mathrm{cm}^{-1}$ was present for all the measurements.

At electric fields of $F_a \sim 15-30\,$mV$\,$cm$^{-1}$, the Stark shift appears as a broadening of the spectral line that increases linearly as a function of $F_a$, Fig.~\ref{fig1}c (see supplementary materials). The broadening increases non-linearly at very small $F_a \sim 10\,$mV$\,$cm$^{-1}$, due to the presence of $F_h \sim 15 \mathrm{\:mV\:cm}^{-1}$. From the broadenings shown in Fig.~\ref{fig1}c, we determined dipole moments for the measured molecular states.  For the $-277 \mathrm{\:MHz}$ vibrational peak near the $37S+6S$ asymptote, the measured electric dipole moment is $D = 2330 \pm 400 \mathrm{\:Debye}$, while we obtained $D = 2310 \pm 250 \mathrm{\:Debye}$ and $D = 1915 \pm 164 \mathrm{\:Debye}$ for the $v=0$ level in the outer wells shown in Fig.~\ref{fig1}b near the $39S+6S$ and $40S+6S$ asymptotes, respectively. The observed PEDMs are $100-1000$ times larger than those measured in previous experiments \cite{li11,bluetrilo} primarily due to the greater degenerate admixture present in these newly observed states. Electron density distributions for each molecular state, $|\Psi(R_e;\mathbf{r}) |^2$, whose PEDM was measured, are shown in Fig.~\ref{fig1}a. In contrast to the Rb results \cite{li11}, it was not necessary to subtract the Cs $nS$-state contributions from $\Psi(R;\mathbf{r})$ to observe the `trilobite'.

The highly localized electron density shown in Fig.~\ref{fig1}a combined with the large internuclear separation, $\sim 120\,$nm, results
in extremely large PEDM, $d = \langle \chi_v (R)\vert d(R) \vert \chi_v (R)\rangle$ where $d(R) = \langle \Psi ( R;\mathbf{r})\vert z \vert \Psi( R;\mathbf{r})\rangle$ is the $R$ dependent dipole moment, and $\vert \chi_v (R)\rangle$ is the $v$-th vibrational wave function. The $R$-dependent dipole moment $d(R)$ is almost entirely due to the hybridization by the high-$l$ degenerate Rydberg manifold. $\Psi(R;\mathbf{r})$ can be written,
\begin{equation}
\Psi(R;\mathbf{r})  = c_{S}(R)  \psi_{nS}(\mathbf{r})  +  c_{T}(R)  \psi_{(n-4)T}(R;\mathbf{r}),
\end{equation}
where the $\psi_{(  n-4) T}(R;\mathbf{r})$ trilobite state is a linear combination of $l>2$ states that maximally localizes the Rydberg electron near the ground state perturber. Here $c_{S}(R)$ and $c_{T}(R)$ are the probability amplitudes for finding the electron in an $nS$-state or trilobite states, respectively. The $R$-dependent dipole moment is $d(R) \cong \vert c_{T}(R)\vert^{2}d_{T}(R)$ where $d_{T}(R)\propto (n-4)^{2}R$ is the dipole moment of the bare trilobite molecule. Far from the avoided crossings, $c_{S}(R)$ is approximately proportional to  $c_{S}(R)  \sim (E_{(n-4) l>2}-E_{nS})/U(R)$. Therefore, the PEDM becomes monotonically smaller with increasing $v$ as the higher-lying vibrational wave functions progressively maximize their probability amplitude at the outer turning points. The calculated PEDMs, shown in Fig. \ref{fig1}b, are within $13$\% of the corresponding experimental values. The main source of error in the theoretical PEDMs is the uncertainty in the position of the vibrational state energy levels.  A shift in the binding energy of ~40 MHz in $\nu = 0$ for Cs(37S+6S)$^3\Sigma$ results in an ~80 Debye change in the PEDM. Taking this uncertainty into account, the experimental and theoretical values agree to within one standard deviation.

Fig. \ref{fig2}a shows a comparison between the experimental spectrum and the PEC correlating to the $37S + 6S$ dissociation limit, with $F_h = 15\,$mV$\,$cm$^{-1}$. By applying the same functional form employed for fitting the Stark shifts, we calculated the theoretical spectra shown in Fig.~\ref{fig3}a. There is significant modulation in the peak strength between even and odd $v$. The modulation of the vibrational state amplitude is most clearly observed for the molecular states correlating to the $37S+6S$ asymptote, as well as near the minima of each of the different PEC wells. The modulation is due to odd parity cancelation in the Frank-Condon factors, akin to the Cooper minima in atomic ionization spectra \cite{Cooper}. The de Broglie wavelength of the Cs atoms is $\lambda_{dB} (T=40 \mu\mathrm{K})\sim 35\,$nm, while the width of the potential wells is $w \sim 5\,$nm. Over this span, the ground state wave function is effectively flat.

%
%

The observed lifetimes of each state are shown in Fig. \ref{fig3}b. The lifetimes are shorter than the Cs($nS$) states. For example, the lifetime of the $40S$ state due to radiative and blackbody decay at $300\,$K is $37 \mathrm{\:\mu s}$ \cite{beterov}. The shorter lifetimes are indicative of mixing states such as $(n-4)\mathrm{F}$ Rydberg state, which has a small, yet nonzero, quantum defect. The lifetime of Cs($36$F) is $\tau_{36F} \sim 18 \mathrm{\:\mu s}$ \cite{ga94}. The fact that the decay of the molecular state is faster than the pure $nS$ Rydberg state is another indication of the degenerate mixing involved in the formation of the hybrid trilobite molecules.

We have observed, and quantitatively described, the formation of homonuclear molecules with giant kilo-Debye PEDM. Due to a small noninteger  quantum defect for Cs($nS$)-states, there is substantial mixing of the $(n-4) l> 2$ degenerate electronic manifold and Cs $nS$ wave functions. This hybridization allows the molecule to acquire a giant kilo-Debye PEDM and yet be excited using standard two-color laser schemes. This is a new class of ultralong-range molecule that is an advantageous cross between a traditional `trilobite' molecule and a low-$l$ ultralong-range Rydberg molecule \cite{gre06,bend09,bluetrilo,Eyler13,Pfau14,Raithel14}. The discovery of the trilobite ultralong-range Rydberg molecule could potentially open opportunities in ultracold chemistry and strongly correlated many-body physics due to the fact that these exotic states have engineered mesoscopic localization and kilo-Debye permanent electric dipole moments.



\bibliographystyle{Science}

\begin{scilastnote}
\item {\bf Acknowledgements}: DB, JY and JPS were supported by the NSF grant (PHY-1205392). Support for HRS came from an NSF grant through ITAMP at the Harvard-Smithsonian Center for Astrophysics. STR would like to thank B. M. Peden and B. L. Johnson for insightful discussions.\end{scilastnote}

%

\clearpage

\section*{Supplementary Materials}

The Rydberg molecules were produced in a $1064 \mathrm{\:nm}$ crossed far-off resonance trap (FORT) loaded from a Zeeman slower-loaded magneto-optical trap (MOT). The experimental setup has been described elsewhere \cite{bluetrilo,photoionization}.

The crossed FORT was created using a 50 W Yb fiber laser at 1064 nm. The laser was focused to a $(1/e^2)$ spot size of $98 \pm 1.0 \mathrm{\:\mu m}$. The beam was passed through the trapping region twice, with the second pass at a $22.5^\mathrm{o}$ angle to the first, resulting in an aspect ratio of approximately 2:1. The beams were not interferometrically stabilized, so a $\lambda/2$ waveplate was used to rotate the polarization of the second pass to prevent interference fringes. The trap depth was $T \sim 5 \mathrm{\:mK}$. The trap frequencies were $2 \pi \times 3.58 \mathrm{\:kHz}$ on the short axis and $2 \pi \times 1.0 \mathrm{\:kHz}$ on the long axis. The crossed FORT was loaded to a peak density of $5 \times 10^{13} \mathrm{\:cm}^{-3}$. The atoms inside the trap were optically pumped into the $6S_{1/2}(F=3)$ hyperfine ground state to avoid collisional losses from the trap. The temperature of the atoms in the FORT was $T= 40 \mathrm{\:\mu K}$.

A two-photon process was used to photoassociate the molecular states in the dipole trap. The first step was generated by a near-infrared photon blue-detuned by 275 MHz from the $6S_{1/2}(F=3) \rightarrow 6P_{3/2}(F'=4)$ transition. The second step of the excitation was generated by a tunable dye laser at $\sim 511 \mathrm{\:nm}$, near a transition to an $n\mathrm{S}_{1/2}$ Rydberg state. The detuning of the first step was chosen to place the frequencies of the hyperfine ghost peaks outside the frequency range of interest; to the red of the atomic Rydberg line when scanning with the second step laser. This required a blue detuning on the first step to move the hyperfine ghosts outside that range.

The lasers were switched and frequency-shifted by acousto-optic modulators (AOMs) and coupled into optical fibers to transport the light to the experiment. The first step laser had $\sim 5 \mathrm{\:mW}$ of power and was collimated to a spot size of $1 \mathrm{\:mm}$. The second step laser had $70 \mathrm{\:mW}$ of power and was focused to a spot size of $44 \mathrm{\:\mu m}$. The second step laser co-propagated with the second pass of the FORT beam. The angle between the first step and second step beams was $112.5^\mathrm{o}$.

 A train of 1000 laser pulses at $2 \mathrm{\:kHz}$ began $20 \mathrm{\:ms}$ after the end of the FORT loading period, lasting for $500 \mathrm{\:ms}$. The first step and second step lasers were pulsed simultaneously. Each laser pulse was $10 \mathrm{\:\mu s}$ long. After a delay of $5 \mathrm{\:\mu s}$ to allow the FORT beam to photoionize excited atoms, the ions were projected onto a microchannel plate (MCP) detector using an electric field pulse with a duration of $500 \mathrm{\:ns}$ and amplitude $F_p = 67 \mathrm{\:V}\:\mathrm{cm}^{-1}$ . The electric field plates were also used to generate background electric fields for the Stark shift measurements. After the pulse train was complete, the second-step laser frequency was incremented and the FORT was reloaded.

The spectra were produced by counting the pulses on the MCP at each frequency step, discriminating $\mathrm{Cs}^+$ and $\mathrm{Cs_2}^+$ ions based on the time of flight. No $\mathrm{Cs_2}^+$ ion signal was detected in the experiment. The trilobite states appear as satellite peaks to the red of the atomic $n\mathrm{S}$ Rydberg line in the $\mathrm{Cs}^+$ signal. Ten scans were averaged for each spectrum. Spectra were taken from $\sim 1 \mathrm{\:GHz}$ red of the $n\mathrm{S}_{1/2}$ line.  The peak molecular signals were $\sim 4\mbox{--}5$ orders of magnitude weaker than the atomic $nS_{1/2}$ Rydberg signal. Two principal reasons for the small signal amplitude, as compared to Rb, are the smaller admixture of $S$-state character in the electronic wave functions, and the fact that the outer potential wells shown in Fig.~\ref{fig1}b are located at internuclear separations that are less than the average separation between the atoms in the gas.

The frequency of the first step laser was monitored using a saturated absorption setup. AOMs were used to shift the light in the saturated absorption setup so that it was resonant with the $6S_{1/2}(F=3) \rightarrow 6P_{3/2}(F'=4)$ transition. Additionally, an electromagnetically-induced transparency setup using a room-temperature Cs vapor cell was used to provide frequency markers for the spectra. The horizontal background electric field, $F_h$ was measured by recording a Stark spectrum of the Cs $126D$ Rydberg state with the applied (vertical) electric field, $F_a$, zeroed.

Lifetimes of the trilobite states were measured by varying the time between excitation and ionization. To prevent photoionization from contaminating the lifetime measurement by ionizing Rydberg atoms early, the FORT was switched off for $40 \mathrm{\:\mu s}$, starting $15 \mathrm{\:\mu s}$ before the excitation laser pulses. The FORT was completely off when the laser excitation pulses occurred. The atoms were field ionized at $F_p= 356 \mathrm{\:V\:cm}^{-1}$. Switching the FORT off increased losses from the trap due to thermal expansion of the atomic cloud during the off period. To mitigate this problem, the pulse train was reduced to 100 pulses in $50 \mathrm{\:ms}$. The experiment was performed by holding both steps of the two-photon transition in resonance with the trilobite state and counting ions during a 100,000 pulse period for each delay. The length of the excitation pulse for the lifetime experiment, $5 \mathrm{\:\mu s}$, caused an uncertainty of $\pm 2.5 \mathrm{\:\mu s}$ in each delay measurement. Tunneling of the system out of the potential wells studied in this paper is not relevant, as the tunneling times are far longer than any of the timescales present in the experiments.

Due to the large spatial extent of these molecules and the fairly short lifetimes, $\sim 10\,\mu$s, the excited molecular states are localized angular wave packets, oriented relative to the external electric field. The orientation is determined by the detuning of the excitation from the field free molecular energy. The spectra that result are the same as those obtained for molecules whose dipole moments are randomly oriented with respect to the electric field.  There is a constant probability density of finding molecules oriented with a given projection of dipole moment along the electric field direction. As a result, the linear Stark shift manifests as a broad flat feature in the absorption spectrum that spans $\Delta\omega=\pm dF$ where $d$ is the molecular dipole moment and $F = \sqrt{F^2_h + F^2_a}$ is the total external field amplitude. Because the laser has a finite linewidth and the final electronic state is metastable, the spectral line has the shape of a step function of width $2dF$ convolved with a Lorentzian. This spectral lineshape was used to characterize the molecular Stark shifts.

The peak widths as a function of applied electric field are shown in Fig. \ref{fig1}c. The fitting function is,
\begin{equation}
\frac{A}{2dF} \left[\mathrm{tan}^{-1}\left( \frac{x-x_0+dF}{\gamma/2}\right) - \mathrm{tan}^{-1}\left( \frac{x-x_0-dF}{\gamma/2}\right)\right],
\end{equation}
where the transition probability $A$, the peak center $x_0$, and the spectral line broadening $dF$ are fit parameters. The parameter $x_{0}$ is the field-free vibrational energy, $\gamma \sim 3 \mathrm{\:MHz}$ is the width of the bare molecular transition, determined by the linewidth of the two-photon transition, and the transition probability $A$ is proportional to the square of the s-wave electron probability amplitude matrix element between the initial and final molecular vibrational states.  Thermal averaging over the initial states creates additional broadening on the order of 1 MHz; this additional broadening is incorporated into the width of the bare molecular transition, $\gamma$.  It is assumed that the initial state is a homogeneous distribution of Cs atoms. 
Taking into account the background field, the line broadening increases with electric field as $d \sqrt{F_a^2+F_h^2}$.

In atomic units the Fermi contact interaction is given by
\begin{equation}
V_{e-atom}(\mathbf{r},\mathbf{R})  =  2\pi A_{s}(k)  \delta (\mathbf{r}-\mathbf{R})  \\ +6\pi A_{p}^{3}(k)  \delta(\mathbf{r}-\mathbf{R}) \overleftarrow{\nabla}\cdot\overrightarrow{\nabla}\nonumber \label{Fermipot}
\end{equation}
where $A_{s}(k)$ and $A_{p}(k)$ are the energy dependent $s$- and $p$-wave scattering lengths of the electron-pertuber atom collision system, respectively, $\mathbf{r}$ is the Rydberg electron position vector from the ionic core, and $\mathbf{R}$ is the position vector of the neutral ground state perturbing atom to the Rydberg core. The electron momentum, $k$, is defined within the semi-classical approximation where $E_{b}=k^{2}/2-1/R$. In this expression, $E_b$ is the binding energy of the isolated Rydberg atom. Here, we used the triplet $s$- and $p$-wave scattering lengths found in \cite{bluetrilo, tn91, scheer98}. Because the experiment is carried out in a FORT, rather than a magnetic trap, singlet Rydberg-ground state molecule states are present. However, the Cs singlet scattering lengths are small \cite{bah01} and these states are not included in the calculations. To account for the spin orbit splitting, the $p$-wave scattering length is defined by
\begin{equation}
A_{p}=\sum_{J=0}^{2}\left[  C_{10,1M_{J}}^{JM_{J}}\right]  ^{2}A_{p,J}%
^{3},\label{Eq:pwavescatt}
\end{equation}
where $C_{L_{1}M_{1},SM_{S}}^{JM_{J}}$ is a Clebsch-Gordan coefficient that couples the angular momentum of the Rydberg electron ($L_{1},M_{1}$) to the total spin of the Rydberg electron-ground state atom system ($S,M_{s}$). We set the $^{3}P_{1}$ resonance position to $8\,$meV \cite{scheer98} and the spin-orbit splitting of the $^{3}P_{0}$ and $^{3}P_{1}$ resonances to the calculated values in Ref. \cite{tn91}. For
the electron to experience the $s$-wave contact interaction in Eq.~\ref{Fermipot}, it must have a non-zero probability to be found along the internuclear axis of the ultralong-range Cs dimer. As a consequence, only states with total angular momentum projection, $M_J=M_S+M_L=0$ along this axis, contribute to the potentials.

In Rb Rydberg molecules \cite{li11}, the resulting electronic states are dominated by the $nS$ Rydberg states and the $s$-wave electron-atom scattering. For Rb, the interaction potential mimics the Rydberg electron density function. In contrast, due to the accidental near-degeneracy present in Cs between the $nS$ and $(n-4) l>2$ Rydberg states, scattering of the Cs Rydberg electron between different Rydberg states must be considered. Recall that the ground state atom scattering center lies off-center from the Rydberg atom.  This means that in an $s$-wave collision, the electron can scatter from one angular momentum state into another. Such multiple scatterings result in highly-localized electronic wave functions, dominated by a nearly degenerate manifold of hydrogenic $\left(  n-4\right) l>2$ states.

In other previously studied Rydberg molecule systems \cite{gds00,li11}, $p$-wave spin-orbit splittings were small, and could be treated as a single pole in the $p$-wave scattering length. Due to the large spin-orbit splitting in Cs, the $p$-wave scattering length in Eq.~\ref{Eq:pwavescatt} leads to multiple sets of narrow avoided crossings. The Clebsch-Gordan coefficients in Eq.~\ref{Eq:pwavescatt} create two different families of potentials, classified with $\left\vert M_{J}\right\vert =0$ and $\left\vert M_{J}\right\vert = 1$. For the potential wells in which the states of interest reside, the $M_J=0$ and $M_J=\pm 1$ projections are nearly degenerate, and not distinguishable in the experiment.

For the calculation of the BO potentials, the electronic wave function is dominated by the $(n-4)(l>2)$ hydrogenic manifold and the $ns$ state, however, a larger basis set is needed to achieve agreement with the experimental results.  For this work, we used a Rydberg orbital basis set that includes all of the Rydberg orbits in-between the $(n\pm 1)S$ Rydberg states including the $(n-3)(l>2)$ nearly degenerate manifold, just above the $(n+1)S$ Rydberg threshold.  Specifically we include the $nS$, $(n\pm1)S$, $nP$, $(n-1)P$, $(n-1)D$, and $(n-2)D$ states as well the $(n-3)(l>2)$, $(n-4)(l>2)$,  and $(n-5)(l>2)$ degenerate manifolds.
For the comparison between theory and experiment, a $25 \mathrm{\:MHz}$ average AC Stark shift of the ground state due to the FORT laser was accounted for by shifting the experimental spectra. The FORT-induced AC Stark shift was measured by comparing the spectral position of each Rydberg state in the MOT, used to load the FORT, with the spectral position of the Rydberg state in the FORT.




\end{document}